\documentclass[aps,prb,twocolumn,showpacs]{revtex4}
\usepackage{bm}
\usepackage{amsmath}
\usepackage{amssymb}
\usepackage{amsfonts}
\usepackage{graphicx}

\begin{document}

\title{Induced current  in the presence of magnetic flux tube of small radius}

\author{Alexander I. Milstein}
\author{Ivan S. Terekhov}
\email[]{A.I.Milstein@inp.nsk.su}
\email[]{I.S.Terekhov@inp.nsk.su}
\affiliation{Budker Institute of Nuclear Physics, 630090
Novosibirsk, Russia}

\date{\today}

\begin{abstract}
The  induced current density, corresponding to the massless Dirac
equation in (2+1) dimensions in a magnetic flux tube of small radius is considered. This problem is important for graphene.  In the case, when an electron can not penetrate the region of nonzero magnetic field, this current is the odd periodical function of the magnetic flux. If the region inside the magnetic tube is not forbidden for penetration of electron, the induced  current is not a periodical function of the magnetic flux. However in the limit $R\to 0$, where $R$ is the radius of magnetic flux tube, this function has the universal form  which is independent of the magnetic field distribution inside the magnetic tube at fixed value of the magnetic flux.
\end{abstract}

\pacs{03.65.Vf, 73.43.Cd, 81.05.Uw}

\maketitle

\section{Introduction}

The Aharonov-Bohm effect predicted in  Ref.\cite{AB} has been investigated in numerous
papers, see review \cite{Peskin}. Due to this quantum effect,
the vector potential affects the electron even in the case when the electron can not penetrate
the region of non-zero magnetic field.  This effect has been investigated within
non-relativistic \cite{AB,Hagen,Khallil_1,Khallil_2} and relativistic
\cite{Gerbert1,Gerbert2,Hagen,Khallil_1,Alford} wave equations.
In quantum field theory, similar effect has been
investigated in Refs.\cite{Alford,Gerbert1,Gerbert2,Deser,Alford_2,Yau,Sitenko1,Sitenko2}.
One of the consequences of the Aharonov-Bohm effect is  scattering of a charged particle in the field of an
infinitesimally thin solenoid. It was shown in Refs.\cite{Gerbert1,Gerbert2}  that  the Dirac
Hamiltonian in the background of the field of infinitesimally thin
solenoid requires a self-adjoint extension. This  extension can be parametrized by one  extension parameter $\theta$.
As a result, the eigen functions of the  Hamiltonian and the corresponding scattering cross section  depend on this parameter.

After recent fabrication of a monolayer graphite (graphene), Ref.\cite{Novoselov1}, new possibilities to study the Aharonov-Bohm
effect have appeared. The single electron dynamics in graphene is
described by the massless Dirac equation in (2+1) dimensions, see Refs.\cite{Wallace,McClure,Semenoff,Gonzalez}.
Using the Green's function of this equation in the field of infinitesimally thin solenoid,  the induced current density  has been investigated analytically in Ref.\cite{Jackiw}. Calculations  have been performed exactly in a magnetic flux tube in the case when an electron can not penetrate the region of nonzero  magnetic field. In the recent paper \cite{Slobodeniuk}, the local density of states in graphene was obtained in the field of infinitesimally thin solenoid.

In the present paper, we derive the induced current density, corresponding to the massless Dirac equation in (2+1) dimensions, in the background of the  magnetic flux tube of finite radius $R$. We investigate in detail the asymptotics $R \longrightarrow 0$ of this current density at fixed magnetic flux through the tube  but for  arbitrary distribution of a magnetic field inside the  tube. We consider two cases. In the first case an electron can not penetrate the region of nonzero magnetic field, and in the second case this is allowed.   Naturally, the result obtained in the first case is independent of the distribution of magnetic field inside the flux tube and coincides with the result of Ref.\cite{Jackiw}. More interestingly, in the second case the induced current density  is also independent of the magnetic field distribution  at $R\longrightarrow 0$ . In another words, we find the explicit value of the extension parameter $\theta$,  Refs.\cite{Gerbert2,Sitenko1,Sitenko2}, and show  that it is independent of the magnetic field distribution.

The article is organized as follows. In Sec. \ref{section2} we derive  the Green's
function of an electron in a magnetic field of the flux tube and obtain
the induced current density at distances $r\gg R$. Calculation of the induced current
density in the first and the second cases of the problem is performed
in Sec. \ref{section3} and  Sec. \ref{section4}, respectively. We conclude with a brief summary of our results.

\section{General discussion}\label{section2}
The induced current density, corresponding to the massless Dirac equation in (2+1) dimensions in a magnetic field,   has
the form, Ref.\cite{Jackiw}:
\begin{eqnarray}\label{InducedCurrent}
\bm j_{ind}(\bm r)=-ie\int_C
\frac{d\epsilon}{2\pi}\mathrm{Tr}\{\bm \sigma {G}(\bm r,\bm
r|\epsilon)\}\, ,
\end{eqnarray}
where the  Green's function ${G}(\bm r,\bm r'|\epsilon)$  obeys the equation
\begin{eqnarray}
\left[\epsilon  -\bm\sigma\cdot\left( \bm p-{e}\bm A(\bm
r)\right)\right]{G}(\bm r,\bm r'|\epsilon)=\delta(\bm r-\bm
r')\,.
\end{eqnarray}
Here $\bm\sigma=(\sigma_x,\sigma_y)$, and $\sigma_i$ are the Pauli
matrices; $\bm p=(p_x,p_y)$ is the momentum operator, $\bm
r=(x,y)$,  $\bm A(\bm r)$ is the vector potential of the magnetic
field,  $e$ is the electron charge, the system of units
$\hbar=c=1$. According to the Feynman rules, the contour $C$ of
integration over $\epsilon$ goes below the real axis in the left
half plane and above the real axis in the right half plane of the
complex $\epsilon$ plane. The magnetic field of the  flux  is $\bm
B(r)=B(r)\bm\nu$, where $\bm\nu$  is the unit vector directed
along z axis,  $B(r)$ is a continuous  function which obeys the
condition $B(r)=0$ for $r>R$. It is convenient to choose the
vector potential in the form
\begin{eqnarray}\label{vector_potential}
\bm A(\bm r)= \frac{\Phi[\bm \nu\times\bm r]}{2\pi r}V(r)\,,
\end{eqnarray}
where $\Phi$ is the total magnetic flux through the tube, and a function $V(r)$ is
\begin{eqnarray}
V(r)=\frac{2\pi}{\Phi r}\int_0^{r}dy\,y\,B(y) \,.
\end{eqnarray}

Using the analytical properties of the Green's function, we deform
the contour of integration over $\epsilon$ in Eq.
(\ref{InducedCurrent}) so that it coincides with the imaginary
axis. Thus it is necessary to find the function ${G}(\bm r,\bm
r'|i\epsilon)$ which we represent as
\begin{eqnarray}\label{G_through_D}
{G}(\bm r,\bm r'|i\epsilon)= \left[i\epsilon
+\bm\sigma\cdot\left( \bm p-{e}\bm A(\bm
r)\right)\right]{D}(\bm r,\bm r'|i\epsilon)\,,
\end{eqnarray}
where $\hat{D}(\bm r,\bm r'|i\epsilon)$ is the Green's function of
the squared Dirac equation
\begin{eqnarray}\label{equation_for_D}
\left[\epsilon^2 + \left( \bm p-{e}\bm A(\bm
r)\right)^2-{e}\sigma_z B(r)\right]{D}(\bm r,\bm
r'|i\epsilon)=-\delta(\bm r-\bm r')\,.\nonumber\\
\end{eqnarray}
Substituting  the function ${D}(\bm r, \bm r'| i\epsilon)$ in the form
\begin{eqnarray}\label{D_partialExpansion}
&&\hspace{-1cm}{D}(\bm r,\bm
r'|i\epsilon)=\nonumber\\
&&\hspace{-1cm}\sum_{m=-\infty}^{\infty}\frac{e^{im(\phi-\phi')}}{2\pi}
\left(
\begin{array}{cc}
{\cal D}^{(+)}_m(r,r'|i\epsilon)&0\\
0&{\cal D}^{(-)}_m(r,r'|i\epsilon)
\end{array}
\right)\,,
\end{eqnarray}
and the $\delta$-function as
\begin{equation}
\delta(\bm r -\bm r')=\frac{\delta(r-r')}{2\pi\sqrt{rr'}}
\sum_{m=-\infty}^{\infty} e^{im(\phi-\phi')}
\end{equation}
to  Eq. (\ref{equation_for_D}), we obtain  the following equations
for the functions ${\cal D}^{(\pm)}_m(r,r'|i\epsilon)$
\begin{widetext}
\begin{eqnarray}
\left[\frac{\partial^2}{\partial r^2}+\frac{1}{r}
\frac{\partial}{\partial r}-\epsilon^2 - \left(\frac{m}{r} -
\gamma V(r) \right)^2+ s_{\pm}\frac{\gamma}{r}\frac{\partial
(rV(r))}{\partial r}\right]{\cal
D}^{(\pm)}_m(r,r'|i\epsilon)=\frac{\delta( r- r')}{\sqrt{r
r'}}\,.\label{EqD+-}
\end{eqnarray}
\end{widetext}
Here $s_{\pm}=\pm 1$, $\gamma=\Phi/\Phi_0$, and $\Phi_0=2\pi /e$
is the elementary magnetic flux. In the region $r,r'>R$,  the
function $V(r)$ is $V(r)=1/r$ so that the term with $s_\pm$ in Eq.
(\ref{EqD+-}) vanishes. As a result, the general solution of Eq.
(\ref{EqD+-}) in this region has the form
\begin{widetext}
\begin{eqnarray}\label{general_solution_D}
{\cal
D}^{(\pm)}_m(r,r'|i\epsilon)=-\left(I_{|\lambda|}(|\epsilon|r_<)
K_{|\lambda|}(|\epsilon|r_>)\right. +\left. \beta^{(\pm)}_m
K_{|\lambda|}(|\epsilon|r) K_{|\lambda|}(|\epsilon|r')\right)\, ,
\quad r,r'>R\,.
\end{eqnarray}
\end{widetext} Here $I_\alpha(x)$ and $K_\alpha(x)$ are the
modified Bessel functions of the first and third kind,
respectively, $r_>=\mathrm{max}(r,r')$, $r_<=\mathrm{min}(r,r')$,
and $\lambda=m-\gamma$. The coefficient in front of the first term
in Eq. (\ref{general_solution_D}) follows from the matching
condition for the functions ${\cal D}^{(\pm)}_m(r,r'|i\epsilon)$
at $r=r'$. The coefficients $\beta^{(\pm)}_m$ depend on the
distribution of the magnetic field in the tube. They can be found
from the matching condition at $r=R$ for the solutions
(\ref{general_solution_D}) and the solutions of  Eq. (\ref{EqD+-})
in the region $r<R$.

Substituting  Eqs. (\ref{G_through_D}) and
(\ref{D_partialExpansion}) to  Eq. (\ref{InducedCurrent}),
 we arrive at the following representation for the induced current density at  $r>R$
\begin{widetext}
\begin{eqnarray}\label{current_D}
\bm j_{ind}(\bm r) = \frac{e[\bm\nu \times\bm r]}{4\pi^2 r}
\sum_{m=-\infty}^\infty \int_{-\infty}^\infty d\epsilon\left[
\frac{\lambda}{r}\left( {\cal D}^{(+)}_m(r,r'|i\epsilon) +{\cal
D}^{(-)}_m(r,r'|i\epsilon)\right)\right.
-\left.\frac{\partial}{\partial r}\left( {\cal
D}^{(+)}_m(r,r'|i\epsilon) -{\cal
D}^{(-)}_m(r,r'|i\epsilon)\right) \right]_{r'=r}\, .
\end{eqnarray}
Then we use the solution (\ref{general_solution_D}) for
${\cal D}^{(\pm)}_m(r,r'|i\epsilon)$ and finally obtain
\begin{eqnarray}\label{current_D1}
\bm j_{ind}(\bm r)&=&\bm j_{ind}^{(1)}(\bm r)+\bm
j_{ind}^{(2)}(\bm r)\,,\label{current_D1}\\
\bm j_{ind}^{(1)}(\bm r) &=& -\frac{e[\bm\nu \times\bm r]}{\pi^2
r^2} \sum_{m=-\infty}^\infty \lambda \int_0^\infty d\epsilon \,
I_{|\lambda|}(|\epsilon|r) K_{|\lambda|}(|\epsilon|r)=
\frac{eNv_F[\bm\nu \times\bm r]}{4\pi r^3}\left( \frac{1}{2}
-|\tilde{\gamma}|\right)^2\tan(\pi\tilde{\gamma})
\,, \label{current_J1}\\
\bm j_{ind}^{(2)}(\bm r) &=& -\frac{e[\bm\nu \times\bm r]}{2\pi^2
r} \sum_{m=-\infty}^\infty \int_0^\infty d\epsilon\
\epsilon\left(\beta_m^{(+)}
K_{|\lambda|+\mu}(|\epsilon|r) - \beta_m^{(-)}
K_{|\lambda|-\mu}(|\epsilon|r)
\right)K_{|\lambda|}(|\epsilon|r) \, ,\label{current_J2}
\end{eqnarray}
\end{widetext}
where $\mu=\mbox{sign}(\lambda)$, the quantity $\tilde{\gamma}$ is the fractional part of
$\gamma$, $|\tilde{\gamma}|<1$. It is defined as $\tilde{\gamma}=\gamma-n$ for $\gamma>0$, and
$\tilde{\gamma}=\gamma+n$ for $\gamma<0$, $n$ is the maximal
integer number less than $|\gamma|$. The first contribution,  $\bm
j_{ind}^{(1)}(\bm r)$, coincides  with the result of
Ref.\cite{Jackiw} for the induced current density in the field of  infinitesimally thin solenoid. This result has been obtained  under assumption that an  electron can not penetrate the region of nonzero  magnetic field.
 The contribution  $\bm j_{ind}^{(1)}(\bm r)$ is a
periodical  function of the magnetic flux,  it depends  on the fractional part of the quantity $\gamma$.
 It is equal to zero  at the integer values of $\gamma$ but also at  half-integer
values of this quantity.  The contribution  $\bm j_{ind}^{(2)}(\bm r)$ depends on
the  radius $R$ and distribution of the magnetic field inside the magnetic tube through the coefficients $\beta_m^{(\pm)}$.
These  coefficients are the functions of $\epsilon$, $R$, $\gamma$, and $m$.

\section{The  case of  pure Aharonov-Bohm effect }\label{section3}

In the  pure Aharonov-Bohm effect, an  electron can not penetrate
the region $r<R$ of nonzero magnetic field. The corresponding
boundary condition is $\bm j_{ind}(\bm R)=0$. In this case
${\cal D}_{ m }^{(+)}(r,r'|i\epsilon)={\cal
D}_{m}^{(-)}(r,r'|i\epsilon)$ and we obtain from Eq.
(\ref{current_D})
 for the induced current density
\begin{eqnarray}\label{current_J3}
\hspace{-0.5cm}\bm j_{ind}(\bm r) &=& \frac{e[\bm\nu \times\bm
r]}{\pi^2 r^2} \sum_{m=-\infty}^\infty \lambda\int_0^\infty
d\epsilon\, {\cal D}^{(+)}_m(r,r'|i\epsilon) \, .
\end{eqnarray}
The  boundary condition gives the following form of the
coefficient $\beta_{m}^{(+)}$ in Eq. (\ref{general_solution_D})
\begin{equation}
\beta_{m}^{(+)}=-I_{|\lambda|}(x)/K_{|\lambda|}(x)\, ,\quad x=|\epsilon|{R}\,.
\end{equation}
At $x\ll 1$, the coefficient $\beta_{m}^{(+)}$ in the leading order in $x$ reads
\begin{equation}
\beta_{m}^{(+)}\approx
-\frac{2^{1-2|\lambda|}|\lambda|}{\Gamma^2(|\lambda|+1)}
x^{2|\lambda|}\, .
\end{equation}
Using this expression, we obtain  the asymptotics of the term $\bm
j_{ind}^{(2)}(\bm r)$  in Eq. (\ref{current_J2}) at large
distances $r\gg R$
\begin{widetext}
\begin{eqnarray}
\bm j_{ind}^{(2)}(\bm r) &=& -\frac{e[\bm\nu \times\bm r]}{2\pi
r^3}\sum_{m=-\infty}^{\infty} \lambda |\lambda|
\frac{\Gamma(|\lambda|+1/2)\Gamma(2|\lambda|+1/2)}{\Gamma^3(|\lambda|+1)}
\left(\frac{{R}}{2r}\right)^{2|\lambda|}\label{current_J3_3}
\, .
\end{eqnarray}
The contribution $\bm j_{ind}^{(2)}(\bm r)$  is equal to zero  at
integer and half-integer values of $\gamma$ due to cancelation
between different terms. For other values of $\gamma$ we introduce
the quantity $\tilde{\lambda}=\tilde{m}-\gamma$, where
$\tilde{m}$  minimizes the value $|\gamma-m|$, and obtain  $\bm
j_{ind}(\bm r)$ in the leading and next-to-leading order in
${R}/r$
\begin{equation}\label{current_J3_2}
\bm j_{ind}(\bm r) = \frac{e[\bm\nu \times\bm r]}{4\pi
r^3}\left[\left( \frac{1}{2}
-|\tilde{\gamma}|\right)^2\tan(\pi\tilde{\gamma})
+2\tilde{\lambda}|\tilde{\lambda}|
\frac{\Gamma(|\tilde{\lambda}|+1/2)
\Gamma(2|\tilde{\lambda}|+1/2)}{\Gamma^3(|\tilde{\lambda}|+1)}
\left(\frac{\tilde{R}}{2r}\right)^{2|\tilde{\lambda}|} \right]\, .
\end{equation}
\end{widetext} The leading term $\bm j_{ind}^{(1)}(\bm r)$ tends to
zero at $r\to\infty$ as $1/r^2$, and the correction $\bm
j_{ind}^{(2)}(\bm r)$ diminishes as
$(\tilde{R}/r)^{2|\tilde{\lambda}|}/r^2$.

\section{The case when an  electron can penetrate the region of nonzero magnetic field}\label{section4}

In this section we assume that an electron can penetrate into the magnetic tube.
Let us start with a simple case of the magnetic field distribution of the form
 $B(r)=B_0\theta(R-r)$, where $\theta(x)$ is the step-function.
 In this case the function $V(r)$ reads
\begin{eqnarray}
V(r)=
\dfrac{r}{R^2}\,\theta(R-r)+\dfrac{1}{r}\,\theta(r-R)\,.
\end{eqnarray}
This function is continuous one at $r=R$ so that the  matching
conditions for the functions ${\cal D}^{(\pm)}_m(r,r'|i\epsilon)$
at $r=R$, following from Eq. (\ref{equation_for_D}), are
\begin{eqnarray}\label{matching}
&&\hspace{-0.5cm}{\cal D}^{(\pm)}_m(R-0,r'|i\epsilon)={\cal
D}^{(\pm)}_m(R+0,r'|i\epsilon)\, ,\nonumber\\
&&\hspace{-0.5cm}\left.\frac{\partial {\cal
D}^{(\pm)}_m(r,r'|i\epsilon)}{\partial
r}\right|_{r=R-0}=\left.\frac{\partial {\cal
D}^{(\pm)}_m(r,r'|i\epsilon)}{\partial r}\right|_{r=R+0}  .
\end{eqnarray}
 The solution of Eq. (\ref{EqD+-}) in the region $r<R$, $r'>R$ has the
form:
\begin{eqnarray}\label{solution_r<R}
{\cal D}^{(\pm)}_m(r,r'|i\epsilon)&=&C_m^{(\pm)}(r')\,r^{|m|}
\exp\left\{
-\frac{|\gamma|r^2}{2R^2}\right\}\nonumber\\
&&\times{}_1F_1\left(a^{(\pm)},|m|+1,
\frac{|\gamma|r^2}{R^2}\right)\,,\\
a^{(\pm)}&=&\frac{\epsilon^2R^2}{4|\gamma|}+\frac{|\gamma||m|-\gamma
m+|\gamma|-s_{\pm}\gamma}{2|\gamma|}\,,\nonumber
\end{eqnarray}
where $C_m^{(\pm)}(r')$ are  some functions which can be found
from the conditions (\ref{matching}), ${}_1F_1\left(a,b,x\right)$
is the confluent Hypergeometric function. Using Eqs.
(\ref{general_solution_D}), (\ref{matching}), and (\ref{solution_r<R}) we obtain  for the coefficients $\beta_m^{(\pm)}$
\begin{eqnarray}\label{beta_explicit}
&&\beta_m^{(\pm)}=-\frac{{\cal F}\left[bI_{|\lambda|}(x) + x
I_{|\lambda|+1}(x)\right] -2|\gamma|{\cal F}^{\prime}I_{|\lambda|}(x)}{
{\cal F}
\left[bK_{|\lambda|}(x) - x
K_{|\lambda|+1}(x)\right] -2|\gamma|{\cal F}^{\prime}K_{|\lambda|}(x)}\, ,\nonumber\\
&&{\cal
F}={}_1F_1\left(a^{(\pm)},|m|+1,|\gamma|\right)\,,\nonumber\\
&&{\cal F}^{\prime}=\dfrac{a^{(\pm)}}{|m|+1}{}_1F_1\left(a^{(\pm)}+1,|m|+2,|\gamma|\right)\,,\nonumber\\
&&b=|\lambda|+|\gamma|-|m|\,,\quad x=|\epsilon|R\,.
\end{eqnarray}
It follows from Eqs. (\ref{beta_explicit}) and (\ref{current_J2}) that the induced current
density is the odd function of the parameter $\gamma$, which is a consequence of the Furry theorem.
Below for simplicity we assume   that  $\gamma>0$.

\begin{figure}[t]
\includegraphics[width=0.35\textwidth]{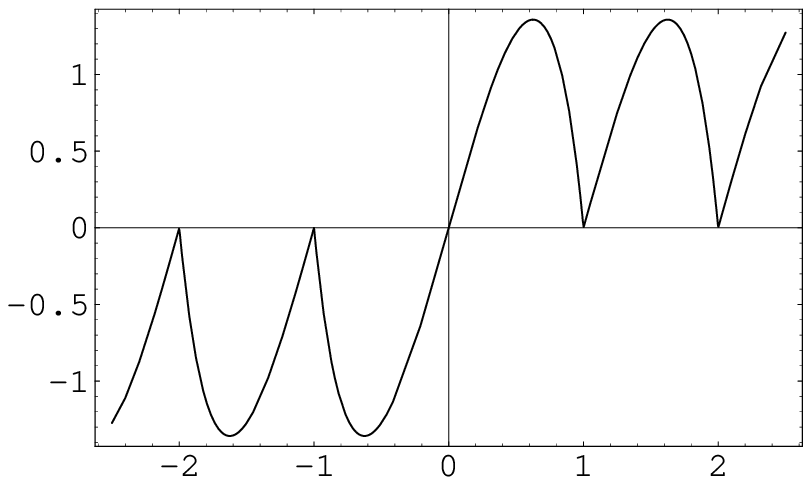}
\caption{Dependence of the function $F(\gamma)$, Eq.
(\ref{currentJ1_f}), on the $\gamma=e\Phi/(2\pi c)$, where $\Phi$
is the magnetic flux.} \label{fig1}
\begin{picture}(0,0)
 \put(80,30){$\gamma$}
 \put(-105,120){$F(\gamma)$}
\end{picture}
\end{figure}

Using Eqs.(\ref{beta_explicit}) and (\ref{current_J2}) it is easy to obtain the asymptotics of the term
 $\bm j_{ind}^{(2)}(\bm r)$ at distances $r\gg R$. In this case, the main contribution to the integral over $\epsilon$
 in Eq. (\ref{current_J2}) is given  by the region $\epsilon\sim 1/r$ so that $x=|\epsilon|R\ll 1$. The main contribution
 to the sum over $m$ in  Eq. (\ref{current_J2}) is given  by
 the term proportional to $\beta_{m}^{(+)}$ with $m=[\gamma]$,  where $[y]$ means the integer part of $y$. For
this $m$ the asymtotics of the coefficient $\beta_m^{(+)}$ at $x\ll 1$ is
\begin{eqnarray}\label{beta_leading}
\beta_m^{(+)}=-\frac{2}{\pi}\sin(\pi\tilde{\gamma})\, .
\end{eqnarray}
 Substituting this expression to Eq. (\ref{current_J2}) and taking the integral over $\epsilon$ we find
\begin{eqnarray}\label{current_J2_rezult}
\bm j_{ind}^{(2)}(\bm r) &=& -\frac{e[\bm\nu \times\bm
r]}{4\pi r^3}\left( \frac{1}{2}
-|\tilde{\gamma}|\right)\tan(\pi\tilde{\gamma})  \, .
\end{eqnarray}
Taking a sum of  $\bm j_{ind}^{(1)}(\bm r)$ and $\bm
j_{ind}^{(2)}(\bm r)$ we finally obtain
\begin{eqnarray}\label{currentJ1_f}
\bm j_{ind}(\bm r)&=& \frac{e[\bm\nu \times\bm r]}{16\pi r^3}
F(\gamma)\,,\nonumber\\
F(\gamma)&=& (1-4\tilde{\gamma}^2)\tan(\pi\tilde{\gamma})\,.
\end{eqnarray}
The function $F(\gamma)$ is shown in Fig.\ref{fig1}.

As it was shown in Ref.\cite{Jackiw}, the induced current in the case of pure Aharonov-Bohm effect is
the periodical function of a magnetic flux. The reason is very simple.
When an electron can not penetrate the region of nonzero magnetic field,
the integer part of $\gamma$ can be eliminated from the wave equation by the gauge transformation.
In opposite case considered in this section, we can not eliminate the integer part of $\gamma$. As a result,
as it is seen from Eq. (\ref{current_J2_rezult}) and  Fig.\ref{fig1}, the induced current density
 is not a periodical function of the magnetic flux.

The result (\ref{currentJ1_f}), obtained for the  magnetic field $B(r)=B_0\theta(R-r)$, coincides with the
corresponding result of Ref.\cite{Sitenko1} for the extension parameter $\theta=\pi/2$.
Now we show that the leading  asymptotics of the induced current density at $R/r\ll 1$ is
independent of the distribution of  magnetic field at the fixed magnetic flux.
As it was pointed out  above, at $R/r\ll 1$ the main contribution to the integral over $\epsilon$
 in Eq. (\ref{current_J2}) is given  by the region $\epsilon\sim 1/r$ so that $|\epsilon|R\ll 1$.
For $r,r'>R$ the solution (\ref{general_solution_D}) is valid for
any distribution of  magnetic field. We use this solution  for
$r'>r\ge R$. For $r'> R\ge r$ we can neglect the term $\epsilon^2$
in Eq. (\ref{EqD+-}) and write the equations for the functions
${\cal D}_m^{(\pm)}(r,r'|i \epsilon)$ in the form:
\begin{eqnarray}\label{newEqForDpm}
&&\hat A_m \hat B_m {\cal D}_m^{(+)}(r,r'|i
\epsilon)=0\,,\nonumber\\
&&\hat B_{m-1}\hat A_{m-1}{\cal D}_m^{(-)}(r,r'|i \epsilon)=0\,,\nonumber\\
&&\hat A_m = \frac{\partial}{\partial r}+\frac{m+1}{r}-\gamma
V(r)\, ,\nonumber \\
&&\hat B_m = \frac{\partial}{\partial r}-\frac{m}{r}+\gamma
V(r)\,.
\end{eqnarray}
The regular at $r=0$ solutions of the Eq. (\ref{newEqForDpm}) have
the form
\begin{widetext}
\begin{equation}\label{D+solution}
{\cal D}^{(+)}_m(r,r'|i\epsilon)= \zeta^{(+)}_m(r')
r^m\exp\left\{-\gamma\int_0^r V(y)dy\right\}\left\{
\begin{array}{lc}
1\, ,& m\ge0\,, \\
\int_0^r x^{-(2m+1)}\exp\left\{2\gamma\int_0^x V(y)dy\right\}dx\,
, &m<0\,,
\end{array}
\right.
\end{equation}
\begin{equation}\label{D-solution}
{\cal D}^{(-)}_m(r,r'|i\epsilon)= \zeta^{(-)}_m(r')
r^{-m}\exp\left\{\gamma\int_0^r V(y)dy\right\}\left\{
\begin{array}{lc}
\int_0^r x^{2m-1}\exp\left\{-2\gamma\int_0^x V(y)dy\right\}dx\, ,& m>0\,, \\
1\, , &m\le0\,.
\end{array}
\right.
\end{equation}
\end{widetext}
Here $\zeta^{(\pm)}_m(r')$ are some functions of $r'$.
Substituting the solutions (\ref{general_solution_D}),
(\ref{D+solution}), and (\ref{D-solution}) to the matching
condition (\ref{matching}) and using the asymptotics of the
functions $I_{\lambda}(x)$ and $K_{\lambda}(x)$ at small $x$ we
obtain the coefficients $\beta_m^{(\pm)}$. In the limit
$|\epsilon|R\to 0$,  only $\beta_{[\gamma]}^{(+)}$ does not vanish
and coincides with the result obtained  for the magnetic field
$B(r)=B_0\theta(R-r)$, see Eq. (\ref{beta_leading}). Thus  the
leading asymptotics of the induced current density is independent
of   the distribution of the magnetic field in the magnetic flux
tube. In our consideration we do not use any assumptions
concerning  the value of  magnetic flux.

\section{Conclusion}\label{section5}
We have investigated the  induced current density, corresponding to the massless Dirac
equation in (2+1) dimensions in a magnetic flux tube of small radius. This problem is important for graphene for which all formulas for the induced current density obtained in our paper should be multiplied by the factor $4 v_F$, the Fermi velocity is $v_F \approx 10^6 \mbox{m/s}\approx c/300$.  In the case of  pure Aharonov-Bohm effect, when an electron can not penetrate the region of nonzero magnetic field, this current is the odd periodical function of the magnetic flux. In the opposite case, when the region inside the magnetic tube is not forbidden for penetration of electron, the induced  current is not a periodical function of the magnetic flux. However in the limit $R\to 0$ this function has the universal form  which is independent of the magnetic field distribution inside the magnetic tube at fixed value of the magnetic flux. Thus it is shown that the extension parameter  $\theta$ introduced in the previous papers has the universal value $\pi/2$. In the same way it is easy to show that
the wave functions corresponding to the massless Dirac equation in (2+1) dimensions in a magnetic flux tube have also the universal form in the limit $R\to 0$.

We are  grateful to R.N.~Lee  for valuable discussions. The work was supported in part by  RFBR
 (grant No 09-02-00024) and the foundation "Dynasty."

\end{document}